\documentclass[12pt]{article}
\usepackage{epsfig}
\usepackage{amstex}
\numberwithin{equation}{section}
\begin{document}

\begin{center}

{\bf\large A COVARIANT AND GAUGE-INVARIANT ANALYSIS OF CMB
ANISOTROPIES FROM SCALAR PERTURBATIONS}\\
\vspace{0.4cm}
{\large Anthony Challinor\footnote{E-mail: A.D.Challinor@@mrao.cam.ac.uk} and Anthony Lasenby\footnote{E-mail: A.N.Lasenby@@mrao.cam.ac.uk}}\\
{\it MRAO, Cavendish Laboratory, Madingley Road,} \\
{\it Cambridge CB3 0HE, UK.}\\
\vspace{0.4cm}
\today
\vspace{0.4cm}

\end{center}
\begin{abstract}
We present a new, fully covariant and manifestly gauge-invariant expression
for the temperature anisotropy in the cosmic microwave background radiation
resulting from scalar perturbations.
We pay particular attention to gauge issues such
as the definition of the temperature perturbation and the placing of the last
scattering surface.
In the instantaneous recombination approximation,
the expression may be integrated up to a Rees-Sciama term for arbitrary matter
descriptions in flat, open and closed universes. We discuss the interpretation
of our result in the baryon-dominated limit using numerical solutions for
conditions on the last scattering surface, and confirm that for adiabatic
perturbations the dominant contribution to the anisotropy on intermediate
scales (the location of the Doppler peaks) may be understood in terms of the
spatial inhomogeneity of the radiation temperature in the baryon rest frame.
Finally, we show how this term enters the usual Sachs-Wolfe type calculations
(it is rarely seen in such analyses) when subtle gauge effects at the last
scattering surface are treated correctly.

\end{abstract}

\section{Introduction}

The calculation of the primary temperature anisotropy in the cosmic
microwave background radiation (CMB) resulting from density perturbations
has a long history, beginning with the seminal paper by Sachs and
Wolfe~\cite{sachs67}. Since the original Sachs-Wolfe estimate, a wealth of
detailed predictions for the anisotropies expected in various cosmological
models have been worked out. The calculations are straightforward in principle,
but, like many topics in cosmological perturbation theory, are plagued
by subtle gauge issues~\cite{stoeger95a}.

The problems of gauge-mode solutions to the linear
perturbation equations and the gauge-ambiguity of initial conditions can
be eliminated by working exclusively with gauge-invariant variables, as in
the widely used Bardeen approach~\cite{bardeen80} and the less well known
covariant approach advocated by Ellis and coworkers~\cite{ellis89a,ellis89b}.
However, gauge issues still arise in connection with the definition of the
temperature perturbation and the
placement of the last scattering surface~\cite{stoeger95a,ellis-er97}.
The latter gauge issues do
not arise at first-order in numerical calculations which integrate the
Boltzmann equation in a perturbed universe, since the visibility function
(which determines the position of the last scattering surface) multiplies
first-order variables giving only a second-order error from the use of a
zero-order approximation to the visibility~\cite{LC-scalcmb}.
However, this is not always the case in Sachs-Wolfe style analyses, which
integrate along null geodesics back to the surface of last scattering,
unless care is taken to ensure that the final
result involves only first-order variables on the last scattering
surface, which then only need be located to zero-order.

In this paper, we present a new expression for the CMB temperature anisotropy
arising from linear scalar perturbations which is fully covariant and
manifestly gauge-invariant. We obtain our expression by integrating the
covariant and gauge-invariant Boltzmann equation~\cite{LC-scalcmb,LC97-er}
along observational null geodesics, paying careful attention to the gauge
issues discussed above. Unlike some covariant results in the
literature (see, for example,~\cite{LC97-er,dunsby96b}), the expression
derived here can be integrated trivially, in the instantaneous recombination
approximation, up to a Rees-Sciama term in
universes with arbitrary matter descriptions. (The covariant results
in~\cite{LC97-er,dunsby96b} can only be integrated in baryon-dominated
universes, thus excluding CDM dominated universes, and other such models
favoured by observation.)
We base our treatment on the physically appealing covariant and gauge-invariant
formulation of perturbation theory, as described in~\cite{ellis89a,ellis89b}.
In this approach, one works exclusively with gauge-invariant variables which
are covariantly-defined and hence physically observable in principle. The
covariant method has many advantages over other gauge-invariant approaches
(such as that formulated by Bardeen~\cite{bardeen80}). Most notably, the
covariant variables have transparent physical definitions which ensures that
predictions are always straightforward to interpret physically. Other
advantages include the unified treatment of scalar, vector and tensor modes,
a systematic linearisation procedure which can be extended to consider
higher-order effects (the covariant variables are exactly gauge-invariant,
independent of any perturbative expansion), and the ability to linearise
about a variety of background models, such as Friedmann-Robertson-Walker (FRW)
or Bianchi models.

For universes which are baryon-dominated at last scattering, our expression
for the temperature anisotropy may be compared to other gauge-invariant
analytic results in the literature. We show that, with suitable approximations,
the result derived here reduces to that given by Panek~\cite{panek86}
and corrects a similar result given by Dunsby recently~\cite{dunsby96b}.
For the baryon-dominated universe, we use
numerical results for the covariant, gauge-invariant variables on the
last scattering surface, obtained from a gauge-invariant Boltzmann
code~\cite{LC-scalcmb},
to discuss the different physical contributions to the primary temperature
anisotropy.
In particular, we show that on intermediate and small scales, the
``monopole'' contribution to the temperature anisotropy is described
by the spatial gradient of the photon energy density, in the energy-frame,
on the last scattering surface. Since the (real) last scattering surface is
approximately a surface of constant
radiation temperature (so that recombination does occur there), the
inhomogeneity of the radiation energy density in the energy-frame
determines a distortion of the last scattering surface relative to the
surfaces of simultaneity in the energy-frame. The extra redshift
(due to the expansion of the universe) which the photons incur due to the
distortion is seen as a ``monopole'' contribution to the temperature
anisotropy on intermediate scales. There is a significant ``dipole''
contribution to the anisotropy on intermediate and small scales, which we
discuss also.

We end with a discussion of the gauge issues inherent in the original
Sachs-Wolfe calculation of the CMB anisotropy~\cite{sachs67}, focusing on the
``monopole'' contribution to the temperature anisotropy on intermediate
scales, described above. This contribution is often missed in Sachs-Wolfe type
calculations through an incorrect treatment of gauge effects at the last
scattering surface~\cite{ellis-er97}. (Equivalently, the term is
often missed through a failure to recognise the direction-dependence of the
``expected temperature'' used to define the temperature perturbation in many
calculations.) This often neglected term, which is not important on large
scales, is an essential component of the Doppler peaks in the CMB power
spectrum.

We employ standard general relativity and use a $(+---)$ metric signature.
Our conventions for the Riemann and Ricci tensors are fixed by
$[\nabla_{a},\nabla_{b}] u^{c} = -{{\mathcal{R}}_{abd}}^{c} u^{d}$, and
${\mathcal{R}}_{ab} \equiv {{\mathcal{R}}_{abc}}^{b}$.
We use units with $c=G=1$ throughout.

\section{Covariant Cosmological Perturbations}

In this section, we summarise the covariant approach to perturbations
in cosmology~\cite{ellis89a,ellis89b}
to establish our notation and conventions. We begin by choosing
a velocity $u^{a}$, which is defined physically in such a manner that if the
universe is exactly FRW the velocity reduces
to that of the fundamental
observers. This property of $u^{a}$ is necessary to ensure gauge-invariance
of the variables defined below.
We refer to the choice of velocity as a frame choice. For most of this paper,
it will not be necessary to make a frame choice.
From the velocity $u^{a}$, we construct a projection
tensor into the space perpendicular to $u^{a}$ (the instantaneous rest space of
observers whose velocity is $u^{a}$):
\begin{equation}
h_{ab} \equiv g_{ab} - u_{a}u_{b},
\end{equation}
where $g_{ab}$ is the metric of spacetime. We use the symmetric tensor $h_{ab}$
to define a spatial covariant derivative ${}^{(3)}\nabla^{a}$ which acting on
a tensor ${T^{b \dots c}}_{d\dots e}$
returns a tensor which is orthogonal to $u^{a}$ on every index:
\begin{equation}
{}^{(3)}\nabla^{a} {T^{b \dots c}}_{d\dots e} \equiv
h_{p}^{a} h_{r}^{b} \dots h_{s}^{c} h_{d}^{t}\dots h_{e}^{u} \nabla^{p}
{T^{r\dots s}}_{t\dots u},
\end{equation}
where $\nabla^{a}$ denotes the usual covariant derivative.

The covariant derivative of the velocity decomposes as
\begin{equation}
\nabla_{a}u_{b} = \varpi_{ab} + \sigma_{ab} +  {\textstyle \frac{1}{3}}
\theta h_{ab} + u_{a} w_{b},
\end{equation}
where $\varpi_{ab} = \varpi_{[ab]}$ is the vorticity, which satisfies
$u^{a} \varpi_{ab}=0$, $\sigma_{ab}=\sigma_{(ab)}$ is the shear, which is
orthogonal to $u^{a}$ and traceless, $\theta\equiv \nabla^{a} u_{a} = 3H$
measures the volume expansion rate ($H$ is the local Hubble parameter), and
$w_{a}\equiv u^{b} \nabla_{b} u_{a}$ is the acceleration. In an exact FRW
universe the vorticity, shear and  acceleration vanish identically.
We regard them as first-order variables (denoted
${\mathcal{O}}(1)$) in an almost FRW universe, so that products of such
variables may be dropped in expressions
in the linearised calculations we consider here. Other first-order variables
may be obtained by taking the spatial gradient of scalar quantities. Such
quantities are gauge-invariant by construction since they vanish identically
in an exact FRW universe. We shall make use of the comoving fractional
spatial gradient of the density $\rho^{(i)}$ of a species $i$,
\begin{equation}
{\mathcal{X}}_{a}^{(i)} \equiv \frac{S}{\rho^{(i)}} {}^{(3)}\nabla_{a}
\rho^{(i)},
\end{equation}
and the comoving spatial gradient of the expansion
\begin{equation}
{\mathcal{Z}}_{a} \equiv S{}^{(3)}\nabla_{a} \theta.
\end{equation}
The scalar $S$ is a local scale factor satisfying
\begin{equation}
\dot{S} \equiv u^{a} \nabla_{a} S = H S, \qquad
{}^{(3)}\nabla^{a} S ={\mathcal{O}}(1),
\end{equation}
which removes the effects of the expansion from the spatial gradients
defined above. The vector ${\mathcal{X}}_{a}^{(i)}$ is a manifestly
covariant and gauge-invariant characterisation of the density inhomogeneity.

The matter stress-energy tensor ${\mathcal{T}}_{ab}$ decomposes with respect to
$u^{a}$ as
\begin{equation}
{\mathcal{T}}_{ab} \equiv \rho u_{a} u_{b} + 2 u_{(a}q_{b)} - p h_{ab} +
\pi_{ab},
\end{equation}
where $\rho\equiv {\mathcal{T}}_{ab} u^{a} u^{b}$ is the density of matter
(measured by a comoving observer), $q_{a} \equiv h_{a}^{b} {\mathcal{T}}_{bc}
u^{c}$ is the energy (or heat) flux and is orthogonal to $u^{a}$,
$p\equiv - h_{ab} {\mathcal{T}}^{ab}
/3$ is the isotropic pressure, and the symmetric traceless tensor
$\pi_{ab} \equiv h_{a}^{c} h_{b}^{d} {\mathcal{T}}_{cd} + p h_{ab}$ is the
anisotropic stress, which is also orthogonal to $u^{a}$.
In an exact FRW universe,
isotropy restricts ${\mathcal{T}}_{ab}$ to perfect-fluid form, so that in an
almost FRW universe the heat flux and isotropic stress may be treated as
first-order variables.

The photons are described by a covariant distribution
function $f^{(\gamma)}(E,e)$, where $E=p^{a}u_{a}$ is the energy of a photon
with momentum $p^{a}$, and $e^{a}$ is unit spacelike vector along the
direction of propagation in the frame defined by $u^{a}$. The photon energy
density $\rho^{(\gamma)}$, the heat flux $q^{(\gamma)}_{a}$ and the
anisotropic stress $\pi^{(\gamma)}_{ab}$ are given by integrals
of the three lowest angular moments of the distribution function:
\begin{align}
\rho^{(\gamma)} & =  \int dE d\Omega\, E^{3} f^{(\gamma)}(E,e) \\
q^{(\gamma)}_{a} & = \int dE d\Omega\, E^{3} f^{(\gamma)}(E,e) e_{a} \\
\pi^{(\gamma)}_{ab} &= \int dE d\Omega\, E^{3} f^{(\gamma)}(E,e) e_{a}e_{b}
+ {\textstyle \frac{1}{3}} \rho^{(\gamma)} h_{ab},
\end{align}
where $d\Omega$ denotes an integration over solid angles. Higher-order
symmetric traceless spatial tensors can be used to characterise higher
moments of the distribution function (see, for example,~\cite{ellis83}),
and are useful in numerical simulations of the
CMB anisotropy~\cite{LC-scalcmb}.
We use the temperature difference from the mean (the full sky average) as
our definition of the temperature anisotropy $\delta_{T}(e)$, so that
\begin{equation}
4\delta_{T}(e) \equiv \frac{4\pi}{\rho^{(\gamma)}} \int dE \, E^{3}
f^{(\gamma)}(E,e) - 1.
\end{equation}
The temperature perturbation $\delta_{T}(e)$ is covariantly defined and
gauge-invariant (it vanishes in an exact FRW universe), and is observable
directly. This should be contrasted with the gauge-dependent temperature
perturbation used by some authors (see Section~\ref{sec_sw} for examples).

The final first-order gauge-invariant variables we require derive from the Weyl
tensor ${\mathcal{W}}_{abcd}$, which vanishes in an exact FRW
universe due to isotropy and homogeneity. The electric and magnetic parts of
the Weyl tensor, denoted by ${\mathcal{E}}_{ab}$ and ${\mathcal{B}}_{ab}$
respectively, are
symmetric traceless tensors, orthogonal to $u^{a}$, which we define by
\begin{align}
{\mathcal{E}}_{ab} &\equiv  u^{c}u^{d} {\mathcal{W}}_{acbd} \\
{\mathcal{B}}_{ab} &\equiv - {\textstyle \frac{1}{2}} u^{c} u^{d}
{\eta_{ac}}^{ef} {\mathcal{W}}_{efbd},
\end{align}
where $\eta_{abcd}$ is the covariant permutation tensor with $\eta_{0123}=
- \sqrt{-g}$.

\subsection{Linearised Perturbation Equations}

Over the epoch of interest, the individual matter constituents of the
universe interact with each other only through gravity, except for the
photons and baryons (including electrons) whose dominant interaction with
each other is via Thomson scattering of photons off free electrons.
The variation of the gauge-invariant temperature
perturbation $\delta_{T}(e)$ along null geodesics is given by the (linearised)
covariant Boltzmann equation~\cite{LC-scalcmb,LC97-er}:
\begin{multline}
\delta_{T}(e)' + \sigma_{T}n_{e} \delta_{T}(e) =
\sigma_{ab} e^{a}e^{b} + w_{a} e^{a} - \left(\frac{1}{3} \theta+
\frac{\rho^{(\gamma)\prime}}{4 \rho^{(\gamma)}}\right) 
\left(1+ 4 \delta_{T}(e)\right)\\
-\sigma_{T} n_{e}\left(v^{(b)}_{a} e^{a} -
\frac{3}{16 \rho^{(\gamma)}} \pi^{(\gamma)}_{ab} e^{a} e^{b} \right),
\label{eq_bol}
\end{multline}
where $n_{e}$ is the free electron density (the effects of thermal motion of
the free electrons is ignored), $\sigma_{T}$ is the Thomson cross
section, $v^{(b)}_{a}$ is the baryon velocity relative to $u_{a}$ ($v^{(b)}_{a}
u^{a} ={\mathcal{O}}(2)$), and a prime denotes differentiation with
respect to the
parameter $\lambda$ along the null geodesic, with
$(u_{a}+e_{a})\nabla^{a} \lambda = 1$. In equation~(\ref{eq_bol}) we have
ignored the effects of polarisation. Including polarisation gives only a small
correction to the collision term in the Boltzmann equation due to the
polarisation dependence of the Thomson cross section.
Equation~(\ref{eq_bol}) is valid for
any type of perturbation (scalar, vector or tensor) and for any value of the
spatial curvature. The evolution of the photon density is given by
\begin{equation}
\dot{\rho}^{(\gamma)} +{\textstyle \frac{4}{3}}  \theta \rho^{(\gamma)} +
{}^{(3)}\nabla^{a} q_{a}^{(\gamma)} = 0,
\label{eq_phot1}
\end{equation}
where an overdot denotes differentiation with respect to proper time along the
integral curves of $u^{a}$ ($\dot{\rho}^{(\gamma)} \equiv u^{a} \nabla_{a}
\rho^{(\gamma)}$). Taking the $l=1$ angular moment of the Boltzmann
equation~(\ref{eq_bol}) gives a propagation equation for the photon heat flux:
\begin{equation}
\dot{q}_{a}^{(\gamma)} + {\textstyle \frac{4}{3}}\theta q_{a}^{(\gamma)} +
{}^{(3)}\nabla^{b} \pi_{ab}^{(\gamma)} + {\textstyle \frac{4}{3}}
 \rho^{(\gamma)} w_{a} - {\textstyle \frac{1}{3}}
{}^{(3)}\nabla_{a} \rho^{(\gamma)} = \sigma_{T} n_{e}
\left({\textstyle \frac{4}{3}} \rho^{(\gamma)} v_{a}^{(b)} - q_{a}^{(\gamma)}
\right).
\label{eq_phot2}
\end{equation}
Taking higher-order moments of Eq.~(\ref{eq_bol}) gives a hierarchy of
equations which are used in the covariant numerical calculations of CMB
anisotropies described in~\cite{LC-scalcmb}.

The electrons and baryons may be approximated by a tightly-coupled ideal fluid
with energy density $\rho^{(b)}$, pressure $p^{(b)}$ in the rest frame of the
fluid which has velocity $u_{a}+v^{(b)}_{a}$. To linear order, the
stress-energy tensor of the baryons is
\begin{equation}
{\mathcal{T}}_{ab}^{(b)} = \rho^{(b)} u_{a} u_{b} + 2 (\rho^{(b)} + p^{(b)})
u_{(a}v^{(b)}_{b)} - p^{(b)} h_{ab},
\end{equation}
which shows that the baryon heat flux is $(\rho^{(b)} + p^{(b)})v^{(b)}_{a}$
in the $u_{a}$ frame. The conservation of photon plus baryon stress-energy
gives the propagation equations for the density
\begin{equation}
\dot{\rho}^{(b)} + (\rho^{(b)} + p^{(b)}) \theta +
(\rho^{(b)} + p^{(b)}){}^{(3)}\nabla^{a} v^{(b)}_{a} = 0,
\end{equation}
and the velocity
\begin{multline}
(\rho^{(b)} + p^{(b)})(\dot{v}^{(b)}_{a} + w_{a}) +
{\textstyle \frac{1}{3}}
(\rho^{(b)} + p^{(b)}) \theta v_{a}^{(b)} + \dot{p}^{(b)}v_{a}^{(b)} -
{}^{(3)}\nabla_{a} p^{(b)}=\\
- \sigma_{T} n_{e} \left({\textstyle \frac{4}{3}}\rho^{(\gamma)}
v_{a}^{(b)} - q_{a}^{(\gamma)} \right).
\end{multline}

In this paper we consider only scalar perturbations. In this case,
the magnetic part of the Weyl tensor ${\mathcal{B}}_{ab}$ and the vorticity 
$\varpi_{ab}$ vanish identically. The electric part of the Weyl tensor
${\mathcal{E}}_{ab}$ and the shear $\sigma_{ab}$ do not vanish, and satisfy the
propagation equations
\begin{align}
\dot{{\mathcal{E}}}_{ab} + \theta {\mathcal{E}}_{ab} + {\textstyle \frac{1}{6}}
\kappa [ 3 (\rho+p)\sigma_{ab} + 3
{}^{(3)}\nabla_{(a}q_{b)} - h_{ab}
{}^{(3)}\nabla^{c} q_{c}
- 3 \dot{\pi}_{ab} - \theta \pi_{ab} ] & = 0 \label{eq_edt}\\
\dot{\sigma}_{ab} + {\textstyle \frac{2}{3}} \theta \sigma_{ab}
-{}^{(3)}\nabla_{(a}w_{b)} + {\textstyle \frac{1}{3}} h_{ab}
{}^{(3)}\nabla^{c} w_{c}
+{\mathcal{E}}_{ab} + {\textstyle \frac{1}{2}} \kappa
\pi_{ab} & = 0,
\label{eq_sigdt}
\end{align}
where $\kappa \equiv 8\pi$, and the constraint equations
\begin{align}
{}^{(3)}\nabla^{b}{\mathcal{E}}_{ab} - {\textstyle \frac{1}{6}} \kappa
[2{}^{(3)}\nabla_{a} \rho +
2 \theta q_{a} + 3{}^{(3)}\nabla^{b} \pi_{ab} ] & =0 \label{eq_econs}\\
{}^{(3)}\nabla^{b} \sigma_{ab} -{\textstyle \frac{2}{3}}
{}^{(3)}\nabla_{a} \theta - \kappa q_{a} & = 0.
\label{eq_sigcons}
\end{align}
The density, pressure, heat flux and anisotropic stress appearing in these
equations are total variables obtained by summing over all matter constituents.

For scalar perturbations, the temporal and spatial aspects of the problem may
be separated by expanding all first-order gauge-invariant variables in tensors
derived from the scalar harmonic functions $Q^{(k)}$, which are defined
covariantly as eigenfunctions of the generalised Helmholtz equation
${}^{(3)}\nabla^{2}Q^{(k)} =
{\textstyle \frac{k^{2}}{S^{2}}}Q^{(k)}$~\cite{haw66}
satisfying $\dot{Q}^{(k)}={\mathcal{O}}(1)$. Specifically, we have
\begin{alignat}{2}
{\mathcal{X}}_{a}^{(i)} & = \sum_{k} k {\mathcal{X}}_{k}^{(i)} Q^{(k)}_{a},
&\qquad {\mathcal{Z}}_{a} & =
\sum_{k} {\textstyle \frac{k^{2}}{S}}{\mathcal{Z}}_{k} Q^{(k)}_{a} \\
q^{(i)}_{a} & = \rho^{(i)} \sum_{k} q_{k}^{(i)} Q^{(k)}_{a} , & \qquad 
\pi^{(i)}_{ab} & = \rho^{(i)} \sum_{k} \pi_{k}^{(i)} Q^{(k)}_{ab}
\label{eq_qpi} \\
{\mathcal{E}}_{ab} & = \sum_{k}{\textstyle \frac{k^{2}}{S^{2}}}\Phi_{k}
Q^{(k)}_{ab}, & \qquad \sigma_{ab}  & = \sum_{k}{\textstyle \frac{k}{S}} 
\sigma_{k} Q^{(k)}_{ab} \\
v^{(b)}_{a}  & = \sum_{k} v_{k}^{(b)} Q^{(k)}_{a} , & \qquad  w_{a} & = 
\sum_{k} w_{k} Q^{(k)}_{a}.
\end{alignat}
The scalar expansion coefficients, such as ${\mathcal{X}}_{k}^{(i)}$, are
themselves first-order gauge-invariant variables, and they satisfy
${}^{(3)}\nabla^{a} {\mathcal{X}}_{k}^{(i)} = {\mathcal{O}}(2)$.
The spatial
vector $Q^{(k)}_{a}$ and the spatial tensor $Q^{(k)}_{ab}$, which is
symmetric and traceless, are defined by
\begin{equation}
Q^{(k)}_{a} \equiv {\textstyle \frac{S}{k}} {}^{(3)}\nabla_{a} Q^{(k)}, \qquad
Q^{(k)}_{ab}\equiv {\textstyle \frac{S^{2}}{k^{2}}}{}^{(3)}
\nabla_{(a}{}^{(3)}\nabla_{b)}Q^{(k)}
- {\textstyle \frac{1}{3}} h_{ab} Q^{(k)}.
\end{equation}
Some useful properties of the scalar harmonics and
derived tensors are summarised in the appendix to
Bruni~{\emph{et al.}~\cite{bruni92}.
This completes the definitions of quantities required in this paper.
Further details of our notation and conventions may be found
in~\cite{LC-scalcmb,LC97-er}.

\section{A Covariant Expression for the Temperature Anisotropy}
\label{sec-kin}

The gauge-invariant CMB temperature anisotropy along a given direction is
obtained by integrating the Boltzmann equation~(\ref{eq_bol}) along the
null geodesic (whose tangent projects onto the given direction) through the
observation point.
Before integrating Eq.~(\ref{eq_bol}), it is convenient to rewrite the
first-order factor multiplying $1+4\delta_{T}(e)$ on the right-hand side in
terms of gauge-invariant variables as follows:
\begin{equation}
\frac{1}{3} \theta + \frac{\rho^{(\gamma)\prime}}{4\rho^{(\gamma)}} =
\frac{1}{4\rho^{(\gamma)}} \left(e_{a}{}^{(3)}\nabla^{a} \rho^{(\gamma)} -
{}^{(3)}\nabla^{a} q_{a}^{(\gamma)}\right),
\end{equation}
where we have made use of the equation of motion of the photon
density, Eq.~(\ref{eq_phot1}), and $(u^{a} + e^{a}){}^{(3)}\nabla_{a}
\lambda = 1$.
At this point, we specialise to scalar perturbations and introduce the
harmonic expansions of the gauge-invariant variables given in the previous
section.
We have
\begin{align}
{\frac{1}{\rho^{(\gamma)}}}{}^{(3)}\nabla^{a} q_{a}^{(\gamma)} & = \sum_{k}
{\textstyle \frac{k}{S}}q_{k}^{(\gamma)} Q^{(k)} \\
{\frac{1}{\rho^{(\gamma)}}}e_{a} {}^{(3)}\nabla^{a}\rho^{(\gamma)} & = 
\sum_{k} {\mathcal{X}}_{k}^{(\gamma)}Q^{(k)\prime},
\end{align}
so that
\begin{equation}
{\frac{1}{4\rho^{(\gamma)}}} \left({}^{(3)}\nabla^{a}q_{a}^{(\gamma)} -
e_{a}{}^{(3)}\nabla^{a}
\rho^{(\gamma)}\right) = -\frac{1}{3} \sum_{k} \left(
{\textstyle \frac{k}{S}} {\mathcal{Z}}_{k}-{\textstyle \frac{S}{k}}
\theta w_{k}\right) Q^{(k)} - \frac{1}{4} \sum_{k}
({\mathcal{X}}_{k}^{(\gamma)}Q^{(k)})',
\end{equation}
where we have used the equation
\begin{equation}
{\textstyle \frac{k}{S}} q_{k}^{(\gamma)} = - {\textstyle \frac{4}{3}}
{\textstyle \frac{k}{S}} {\mathcal{Z}}_{k} +
{\textstyle \frac{4}{3}}{\textstyle \frac{S}{k}} \theta w_{k} -
{\mathcal{X}}_{k}^{(\gamma)\prime},
\end{equation}
which follows from taking the spatial gradient of Eq.~(\ref{eq_phot1}) and
harmonically expanding the result,
to eliminate $q_{k}^{(\gamma)}$ in favour of ${\mathcal{Z}}_{k}$ and the
acceleration. Integrating the
Boltzmann equation~(\ref{eq_bol}) along the null geodesic connecting the
reception point $R$ (where $\lambda=\lambda_{R}$) and a point in the
distant past (where $\lambda=\lambda_{i}$), we find
\begin{multline}
\left(\delta_{T}(e)\right)_{R} = -\frac{1}{4} \sum_{k}
\left({\mathcal{X}}_{k}^{(\gamma)}Q^{(k)} \right)_{R} \\
+ \sum_{k}\int_{\lambda_{i}}^{\lambda_{R}} 
e^{\mbox{\footnotesize $-\tau$}}
\left[{\textstyle \frac{k}{S}}  \sigma_{k} e^{a}e^{b}Q^{(k)}_{ab} -
{\textstyle \frac{1}{3}}
\left({\textstyle \frac{k}{S}} {\mathcal{Z}}_{k} - {\textstyle \frac{S}{k}}
\theta w_{k} \right)
Q^{(k)} + w_{k} e^{a} Q^{(k)}_{a} \right] \, d\lambda \\
+ \sum_{k} \int_{\lambda_{i}}^{\lambda_{R}} - \tau'
e^{\mbox{\footnotesize $-\tau$}}
\left[{\textstyle \frac{3}{16}} \pi_{k}^{(\gamma)} e^{a}e^{b}Q^{(k)}_{ab} -
v_{k}^{(b)} e^{a}Q^{(k)}_{a} +
{\textstyle \frac{1}{4}}{\mathcal{X}}_{k}^{(\gamma)}Q^{(k)} \right] \,
d\lambda, \label{eq_deltat1}
\end{multline}
where $(M)_{R}$ denotes the value of the quantity $M$ evaluated at the point
$R$, and $\tau(\lambda)$ is the optical depth along the line of sight, defined
by
\begin{equation}
\tau(\lambda) \equiv \int_{\lambda}^{\lambda_{R}} n_{e} \sigma_{T} \, d\lambda.
\end{equation}
On angular scales larger than $8'$ we may approximate the
visibility function $-\tau'e^{\mbox{\footnotesize $-\tau$}}$
by a delta function whose support
defines the last scattering surface (the instantaneous recombination
approximation). With this approximation, Eq.~(\ref{eq_deltat1}) integrates to
\begin{multline}
\left(\delta_{T}(e)\right)_{R} = \sum_{k} \left(
{\textstyle \frac{1}{4}}
{\mathcal{X}}_{k}^{(\gamma)}Q^{(k)} + {\textstyle \frac{3}{16}}
\pi_{k}^{(\gamma)} e^{a}e^{b}Q^{(k)}_{ab}
- v_{k}^{(b)} e^{a}Q^{(k)}_{a} \right)_{A} \\
+ \sum_{k} \int_{\lambda_{A}}^{\lambda_{R}} \bigl\{
{\textstyle \frac{k}{S}} \left[
\sigma_{k} \left( {\textstyle \frac{S}{k^{2}}} (SQ^{(k)\prime})' +
{\textstyle \frac{1}{3}} Q^{(k)}\right) - {\textstyle \frac{1}{3}} 
{\mathcal{Z}}_{k} Q^{(k)}
\right] \\
+ {\textstyle \frac{S}{k}} \left( w_{k}Q^{(k)\prime} +
H w_{k}Q^{(k)}\right) \bigr\} \,d\lambda,
\label{eq_deltat2}
\end{multline}
where the point $A$ (where $\lambda=\lambda_{A}$) is the point of intersection
of the null geodesic with the last scattering surface, and we have used the
result
\begin{equation}
e^{a}e^{b}Q^{(k)}_{ab} = {\textstyle \frac{S}{k^{2}}} (SQ^{(k)\prime} )' +
{\textstyle \frac{1}{3}}Q^{(k)}.
\end{equation}
We have dropped a direction independent (monopole) term evaluated at $R$
from Eq.~(\ref{eq_deltat2}) since it will eventually be cancelled by other
monopole terms in the integral. Note that the integrand in
Eq.~(\ref{eq_deltat2}) contains
only kinematic gauge-invariant variables (the shear, the spatial gradient
of the expansion $\theta$ and the acceleration), which simplifies the next
stage in the integration. In~\cite{LC97-er} we gave a more general
expression for the
anisotropy, valid for all perturbation types, but the integrand involved the
spatial gradient of the baryon
density which could only be replaced by the spatial gradient of the total
density if the universe is baryon dominated at recombination. The
expression~(\ref{eq_deltat2}) proves to be
more convenient for the discussion of CMB anisotropies in multicomponent
universes where only scalar perturbations are present.

Integrating the last term in Eq.~(\ref{eq_deltat2}) by parts twice, we find
that
\begin{multline}
\int_{\lambda_{A}}^{\lambda_{R}}{\textstyle \frac{k}{S}}  \left[
\sigma_{k} \left( {\textstyle \frac{S}{k^{2}}} (SQ^{(k)\prime})' +
{\textstyle \frac{1}{3}}
Q^{(k)}\right) - {\textstyle \frac{1}{3}} {\mathcal{Z}}_{k}
Q^{(k)} \right]
+ {\textstyle \frac{S}{k}} \left[ w_{k}Q^{(k)\prime}  + H w_{k}
Q^{(k)}\right] \, d\lambda  \\
=\left[ \sigma_{k} e^{a}Q^{(k)}_{a} -
{\textstyle \frac{S}{k}} (\dot{\sigma}_{k}-w_{k})Q^{(k)} \right]_{A}^{R}
\\
+ \int_{\lambda_{A}}^{\lambda_{R}} \left[
\left({\textstyle \frac{S}{k}}  \sigma_{k}'\right)'
+ {\textstyle \frac{1}{3}} {\textstyle \frac{k}{S}} (\sigma_{k} -
{\mathcal{Z}}_{k}) -{\textstyle \frac{S}{k}} w_{k}' \right]Q^{(k)}\, d\lambda. 
\label{eq_intpart}
\end{multline}
The integrand on the right-hand side of Eq.~(\ref{eq_intpart}) may be
simplified by using the linearised propagation and constraint equations
for the shear and the electric part of the Weyl tensor.
The harmonic expansions of equations~(\ref{eq_edt}) and~(\ref{eq_sigdt})
give
\begin{align}
\left({\textstyle \frac{k}{S}}\right)^{2}(\dot{\Phi}_{k} +
{\textstyle \frac{1}{3}} \theta \Phi_{k})
+ {\textstyle \frac{1}{2}}{\textstyle \frac{k}{S}}  \kappa\rho
(\gamma \sigma_{k} +
q_{k}) +{\textstyle \frac{1}{6}} \kappa \rho \theta (3\gamma -1) \pi_{k} -
{\textstyle \frac{1}{2}} \kappa \rho
\dot{\pi}_{k} & = 0 \label{eq_edthar}
\\
{\textstyle \frac{k}{S}} (\dot{\sigma}_{k} + {\textstyle \frac{1}{3}}
\theta \sigma_{k}-w_{k} ) + \left(
{\textstyle \frac{k}{S}}\right)^{2} \Phi_{k} + {\textstyle \frac{1}{2}}
\kappa \rho \pi_{k} & =  0,
\label{eq_sigdthar}
\end{align}
where $\gamma$ is defined by $p=(\gamma-1)\rho$,
and the constraint equation~(\ref{eq_sigcons}) gives
\begin{equation}
{\textstyle \frac{2}{3}} \left({\textstyle \frac{k}{S}}\right)^{2}
\left[{\mathcal{Z}}_{k} - \left(1-{\textstyle \frac{3K}{k^{2}}}\right)
\sigma_{k}\right] + \kappa \rho q_{k} = 0.
\label{eq_sigconshar}
\end{equation}
In these equations, the scalar variables $q_{k}$ and $\pi_{k}$ are the harmonic
expansion coefficients of the total heat flux and anisotropic stress. They are
related to the component variables $q_{k}^{(i)}$ and $\pi_{k}^{(i)}$
(defined by Eq.~(\ref{eq_qpi})) by
\begin{equation}
\rho q_{k} = \sum_{i} \rho^{(i)} q_{k}^{(i)}, \qquad
\rho \pi_{k} = \sum_{i} \rho^{(i)} \pi_{k}^{(i)},
\end{equation}
where the sums are over individual components $i$.

Evaluating the integrand in Eq.~(\ref{eq_intpart}) by differentiating
the shear propagation equation~(\ref{eq_sigdthar}), substituting for
$q_{k}$ and ${\mathcal{Z}}_{k}$ from equations~(\ref{eq_edthar})
and~(\ref{eq_sigconshar}), and using the zero-order Friedmann equation
\begin{equation}
H^{2} + {\textstyle \frac{K}{S^{2}}} ={\textstyle \frac{1}{3}}  \kappa \rho,
\end{equation}
we find the result
\begin{equation}
\left({\textstyle \frac{S}{k}} \sigma_{k}'\right)' + {\textstyle \frac{1}{3}}
{\textstyle \frac{k}{S}} (\sigma_{k}
-{\mathcal{Z}}_{k}) - {\textstyle \frac{S}{k}} w_{k}' = -2 \dot{\Phi}_{k},
\label{ani_eq_16}
\end{equation}
which is true for scalar perturbations, independent of the
matter description and spatial curvature.
With this, we obtain our final result for the temperature
anisotropy (which is exact in linear theory on angular scales where
instantaneous recombination is valid):
\begin{multline}
\left(\delta_{T}(e)\right)_{R} = \sum_{k} \left( \left[
{\textstyle \frac{1}{4}}
{\mathcal{X}}_{k}^{(\gamma)} +{\textstyle \frac{S}{k}}
(\dot{\sigma}_{k}-w_{k}) \right]Q^{(k)}
\right)_{A} - \sum_{k} \left([v_{k}^{(b)} + \sigma_{k}]
e^{a}Q^{(k)}_{a}\right)_{A}\\
+ {\textstyle \frac{3}{16}} \sum_{k} \left(\pi_{k}^{(\gamma)}
e^{a}e^{b}Q^{(k)}_{ab}\right)_{A}
- 2 \sum_{k} \int_{\lambda_{A}}^{\lambda_{R}}
\dot{\Phi}_{k}Q^{(k)} \, d\lambda,
\label{eq_deltat}
\end{multline}
where we have dropped a (frame-dependent) dipole term evaluated at $R$ since
such a term cannot be distinguished from a first-order peculiar velocity
of the observer at $R$. The final term in Eq.~(\ref{eq_deltat}) describes the
Rees-Sciama effect, which only makes a small contribution to
the anisotropy in $K=0$ models that are matter dominated at recombination
(for $K=0$, $\Phi_{k}$ is approximately constant while a mode is outside the
horizon and during the matter dominated era on all scales).
The third term on the right-hand side of Eq.~(\ref{eq_deltat})
represents a small contribution to the anisotropy from photon anisotropic
stress at last scattering. The sum of the first and second terms
dominates the CMB anisotropy in a $K=0$ universe, with the
relative importance of each term being dependent on $\Omega_{b}$ and $H_{0}$.
Expression~(\ref{eq_deltat}) is a generalisation of the result given by
Dunsby in Section 5 of~\cite{dunsby96b} which was valid only for universes
that are fully baryon dominated at last scattering and are spatially flat.
We shall see in Section~\ref{sec_bar}
how the result in~\cite{dunsby96b} (actually a corrected version of it)
may be obtained from Eq.~(\ref{eq_deltat}) in the limit of baryon domination
at recombination. A similar result to Eq.~(\ref{eq_deltat}) is derived
in~\cite{hu95b} in terms of Bardeen's gauge-invariant variables.

In deriving Eq.~(\ref{eq_deltat}), we have not made an explicit choice for the
velocity $u^{a}$. Each of the four terms on the right-hand side
is frame-independent, which follows from the fact that under a change of
frame $u_{a}\mapsto u_{a} + v_{a}$, where $v_{a}$ is a first-order relative
velocity ($u^{a}v_{a} = {\mathcal{O}}(2)$), ${\mathcal{E}}_{ab}$,
$\dot{{\mathcal{E}}}_{ab}$ and
$\pi_{ab}$ are invariant, while $v_{a}^{(b)} \mapsto v_{a}^{(b)} - v_{a}$
and
\begin{equation}
\sigma_{ab} \mapsto \sigma_{ab} + {}^{(3)}\nabla_{(a}v_{b)} -
{\textstyle \frac{1}{3}} h_{ab}
{}^{(3)}\nabla^{c} v_{c}.
\end{equation}
The frame-invariance of the right-hand side of Eq.~(\ref{eq_deltat})
is necessary since $\delta_{T}(e)$ is invariant in linear theory, up to
the dipole terms that we have dropped from Eq.~(\ref{eq_deltat}).

The non-integral terms on the right-hand side of Eq.~(\ref{eq_deltat}) are
evaluated at the point $A$ on the last scattering surface, which lies on a
null geodesic through the observation point $R$. However, it is only necessary
to locate $A$ to zero-order since the displacement from the ``true'' position
is first-order, which leads to only a second-order error when evaluating
a first-order variable. We shall return to this point in Section~\ref{sec_sw}
where we discuss some of the gauge issues associated with the placement of the
last-scattering surface in the standard calculations of the Sachs-Wolfe effect.

\section{CMB Anisotropy in a Universe Dominated by Baryons at Recombination}
\label{sec_bar}

In cosmological models that are baryon dominated at recombination, the
covariant result for the temperature anisotropy, Eq.~(\ref{eq_deltat}), may be
cast in a more familiar form, which aids direct comparison with other
such results in the literature (for example~\cite{dunsby96b}) and physical
interpretation.

Using the propagation equations~(\ref{eq_sigdthar}) and~(\ref{eq_edthar})
for the shear and the electric part of the Weyl tensor, and the harmonic
expansion of the constraint~(\ref{eq_econs}):
\begin{equation}
2 \left({\textstyle \frac{k}{S}}\right)^{3}\left(1-
{\textstyle \frac{3K}{k^{2}}}\right) \Phi_{k}
- {\textstyle \frac{k}{S}} \kappa \rho \left({\mathcal{X}}_{k} +
\left[1-{\textstyle \frac{3K}{k^{2}}}\right]\pi_{k}\right) -
\kappa \rho \theta q_{k}
=0,
\label{eq_econshar}
\end{equation}
we find in the limit
\begin{equation}
\rho \rightarrow \rho^{(b)}, \quad p \rightarrow 0,\quad {\mathcal{X}}_{k}
\rightarrow {\mathcal{X}}_{k}^{(b)}, \quad q_{k} \rightarrow v_{k}^{(b)}, \quad
\pi_{k} \rightarrow 0,
\end{equation}
at last scattering that
\begin{equation}
{\textstyle \frac{1}{4}} {\mathcal{X}}_{k}^{(\gamma)} +
{\textstyle \frac{S}{k}} (\dot{\sigma}_{k}-w_{k})
\rightarrow {\textstyle \frac{1}{4}} {\mathcal{X}}_{k}^{(\gamma)} -
{\textstyle \frac{1}{3}}{\mathcal{X}}_{k}^{(b)}
- {\textstyle \frac{1}{3}} \Phi_{k} -
{\textstyle \frac{2(6K-k^{2})}{3\kappa\rho S^{2}}}
\Phi_{k} + {\textstyle \frac{2}{\kappa\rho}} H \dot{\Phi}_{k},
\label{eq_monapprox}
\end{equation}
where we have added and subtracted ${\mathcal{X}}_{k}^{(b)}/3$ making use
of Eq.~(\ref{eq_econshar}), and
\begin{equation}
-(\sigma_{k} + v_{k}^{(b)}) \rightarrow
{\textstyle \frac{2k}{\kappa\rho S}}
\left( \dot{\Phi}_{k} + H \Phi_{k}\right).
\end{equation}
It follows that in a universe which is baryon dominated at last scattering,
the temperature anisotropy from scalar perturbations in the instantaneous
recombination approximation becomes
\begin{multline}
\left(\delta_{T}(e)\right)_{R} = \sum_{k} \left( \left[
{\textstyle \frac{1}{4}}
{\mathcal{X}}_{k}^{(\gamma)} - {\textstyle \frac{1}{3}}
{\mathcal{X}}_{k}^{(b)} - {\textstyle \frac{1}{3}}
\Phi_{k} -
{\textstyle \frac{2(6K-k^{2})}{3\kappa\rho S^{2}}} \Phi_{k} +
{\textstyle \frac{2}{\kappa\rho}} H \dot{\Phi}_{k} \right]Q^{(k)}  \right)_{A}
\\
+ \sum_{k} \left( {\textstyle \frac{2k}{\kappa\rho S}}
\left[\dot{\Phi}_{k} + H \Phi_{k}
\right] e^{a}Q^{(k)}_{a}\right)_{A} - 2 \sum_{k}
\int_{\lambda_{A}}^{\lambda_{R}} \dot{\Phi}_{k} Q^{(k)} \, d\lambda.
\label{eq_deltatbarK}
\end{multline}
The first set of terms in square brackets on the right-hand side
of Eq.~(\ref{eq_deltatbarK}) give the ``monopole'' contribution (at last
scattering) to the temperature anisotropy. The term
$({\mathcal{X}}_{k}^{(\gamma)}/4 - {\mathcal{X}}_{k}^{(b)}/3)Q^{(k)}$
arises from entropy perturbations, which may be
characterised covariantly by a vector ${\mathcal{S}}_{a}^{(\gamma b)}$
where
\begin{equation}
{\mathcal{S}}_{a}^{(\gamma b)} \equiv {\frac{3}{4\rho^{(\gamma)}}}
{}^{(3)}\nabla_{a}
\rho^{(\gamma)} - {\frac{1}{\rho^{(b)}}} {}^{(3)}\nabla_{a} \rho^{(b)}.
\end{equation}
For adiabatic initial conditions, entropy
perturbations vanish at last scattering on large scales, where tight-coupling
between the baryons and photons still holds. The second ``monopole'' term,
$-\Phi_{k}Q^{(k)}/3$, is the usual Sachs-Wolfe contribution to the temperature
anisotropy~\cite{sachs67}, whose effect is modified on small scales by the
third and fourth ``monopole'' terms. As noted by Ellis and
Dunsby~\cite{ellis-er97}, the third term is rarely seen in analytic
calculations of the Sachs-Wolfe
effect, although it is present in Panek's result~\cite{panek86}. The
omission arises from subtle gauge effects at the last scattering surface which
we discuss in Section~\ref{sec_sw}. The final monopole term is a small
correction arising from the non-stationarity of the potential $\Phi_{k}$.
The terms under the second summation in Eq.~(\ref{eq_deltatbarK}) make a
``dipole'' contribution to the temperature anisotropy, and are important
on small angular scales.

In a $K=0$ universe, Eq.~(\ref{eq_deltatbarK}) may be written in the form
\begin{multline}
\left(\delta_{T}(e)\right)_{R} = \sum_{k} \left( \left[
{\textstyle \frac{1}{4}}
{\mathcal{X}}_{k}^{(\gamma)} -{\textstyle \frac{1}{3}}
{\mathcal{X}}_{k}^{(b)} -
{\textstyle \frac{1}{3}}\Phi_{k}
+ {\textstyle \frac{2}{3}}H^{-1} \dot{\Phi}_{k} +
{\textstyle \frac{2}{9}} {\mathcal{H}}_{k}^{-2} \Phi_{k}
\right]Q^{(k)} \right)_{A}\\
+ \sum_{k} \left({\textstyle \frac{2}{3}} {\mathcal{H}}_{k}^{-1}
\left[ \Phi_{k}
+ H^{-1} \dot{\Phi}_{k} \right] e^{a} Q^{(k)}_{a} \right)_{A} - 2 \sum_{k}
\int_{\lambda_{A}}^{\lambda_{R}} \dot{\Phi}_{k} Q^{(k)} \, d\lambda,
\label{eq_deltatbar}
\end{multline}
where ${\mathcal{H}}_{k}\equiv SH/k$ is the ratio of proper wavelength to the
Hubble radius. This result corrects that given by Dunsby in
Section 5 of~\cite{dunsby96b} (his equation (61); note also the difference
in metric signature from that adopted here). Note, in particular, that the
dominant contribution to the CMB anisotropy from adiabatic perturbations
on large scales is $-\Phi_{k}/3$, which is a factor of $3$ smaller than
the result in~\cite{dunsby96b}.

\begin{figure}[t!]
\begin{center}
\epsfig{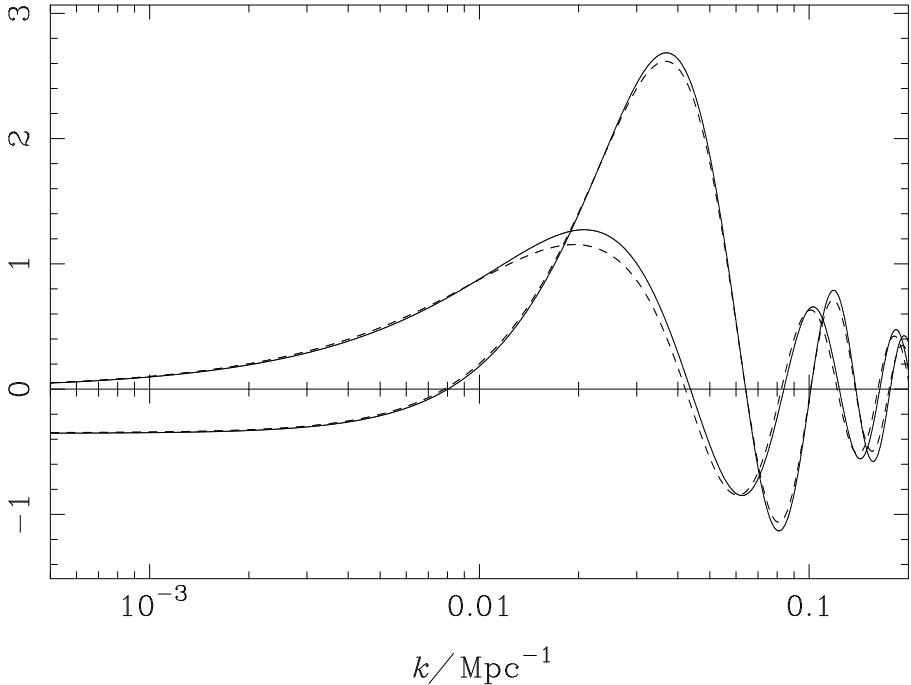}
\end{center}
\caption{{\small Contributions to the CMB temperature anisotropy from
the conditions on the last scattering surface ($z\simeq 1050$)
in a $K=0$ universe with $\Omega_{b}=1$, and
$H_{0}=50\mbox{$\mathrm{kms^{-1}\,Mpc^{-1}}$}$. Only
adiabatic scalar perturbations are considered, and $\Phi_{k}$ is
independent of $k$ initially (so that we are plotting transfer functions).
The curves should be multiplied by the initial $\Phi_{k}$ to give the
actual conditions on the last scattering surface.
The solid lines are calculated from Eq.~(\ref{eq_deltat}); the ``monopole''
contribution (negative on large scales) is
${\mathcal{X}}^{(\gamma)}_{k} /4+ S(\dot{\sigma}_{k} -w_{k})/k$, and
the ``dipole'' contribution (positive on large scales)
is $-\sigma_{k} - v^{(b)}_{k}$. The dashed lines are given by the
approximate expression~(\ref{eq_deltatbar}); the ``monopole'' contribution
(negative on large scales) is ${\mathcal{X}}_{k}^{(\gamma)}/4 -
{\mathcal{X}}_{k}^{(b)}/3-\Phi_{k}/3 + 2H^{-1}\dot{\Phi}_{k}/3 +
2{\mathcal{H}}_{k}^{-2} \Phi_{k}/9$,
and the ``dipole'' contribution
is $2{\mathcal{H}}_{k}^{-1}(\Phi_{k} + H^{-1}\dot{\Phi}_{k})/3$.
The small discrepancy between the solid and dashed curves
is due to the small but non-zero values of
$\rho^{(\gamma)}/\rho$ and $\rho^{(\nu)}/\rho$
at last scattering.}}
\label{fig_com}
\end{figure}

In Fig.~\ref{fig_com} we plot the first two summands in
equations~(\ref{eq_deltat}) and~(\ref{eq_deltatbar}) as a function of $k$,
with $S=1$ at the present, in a $K=0$ universe, in the limit $\Omega_{b}=1$,
where $\Omega_{b}$ is the present-day baryon density in units of the
critical density.
We take $H_{0}=50 \mbox{$\mathrm{kms^{-1}\,Mpc^{-1}}$}$
and consider adiabatic perturbations with only the
fastest growing mode present. At early times, we take $\Phi_{k}$ to be
independent of $k$. The actual conditions on the last
scattering surface are obtained from the transfer functions of
Fig.~\ref{fig_com} by multiplying by the initial values of $\Phi_{k}$
(which are Gaussian random variables in most inflationary theories).
The gauge-invariant variables on the last scattering surface are obtained from
an accurate Boltzmann code employing covariant, gauge-invariant variables, with
adiabatic initial conditions~\cite{LC-scalcmb}. The agreement between the
approximate expression~(\ref{eq_deltatbar}) and the ``exact''
expression~(\ref{eq_deltat}) is good over the full range of $k$ depicted.
The small discrepancy between the sets of curves is due to the small but
non-zero values of $\rho^{(\gamma)}/\rho$ and $\rho^{(\nu)}/\rho$ on the last
scattering surface (which is located at $z\simeq 1050$ in this model).

On large angular scales (small $k$), the ``monopole'' terms dominate the
CMB anisotropy, giving the familiar Sachs-Wolfe plateau for a scale-invariant
spectrum of initial conditions. On smaller angular scales the ``monopole''
term oscillates in $k$ due to the coherent acoustic oscillations in the
photon-baryon plasma that occur for modes inside the (sound) horizon.
These oscillations, along with the oscillations in the ``dipole'' on small
scales, determine the structure of the Doppler peaks in the CMB power spectrum.
For adiabatic perturbations in a $K=0$ universe, the first zero in the
``monopole'' contribution to the anisotropy (the initially lower curves
in Fig.~\ref{fig_com}) occurs where ${\mathcal{H}}_{k} \approx \surd(2/3)$.
This follows from Eq.~(\ref{eq_deltatbar}), the fact that initially adiabatic
perturbations are still adiabatic at last scattering on such scales (see
Fig.~\ref{fig_split}), and the
approximate stationarity of the potential $\Phi_{k}$ in the matter
dominated era of a $K=0$ universe. This effect was noted recently by
Ellis and Dunsby~\cite{ellis-er97}, and was also discussed by
Hu and Sugiyama~\cite{hu95b}. Ellis and Dunsby suggested that this zero
in the ``monopole'' contribution should be observable as a zero in the CMB
power spectrum on angular scales $\simeq 50'$. However, as noted by Hu and
Sugiyama and evident in Fig.~\ref{fig_com}, the ``dipole'' contribution
is already significant on these angular scales, with its first maximum
occurring close to the zero in the ``monopole''. This tends to wash out the
zero (actually a dip once the $k$ modes are summed over) in the power
spectrum, as shown by Hu and Sugiyama~\cite{hu95b} in their Fig. 4. 

\begin{figure}[t!]
\begin{center}
\epsfig{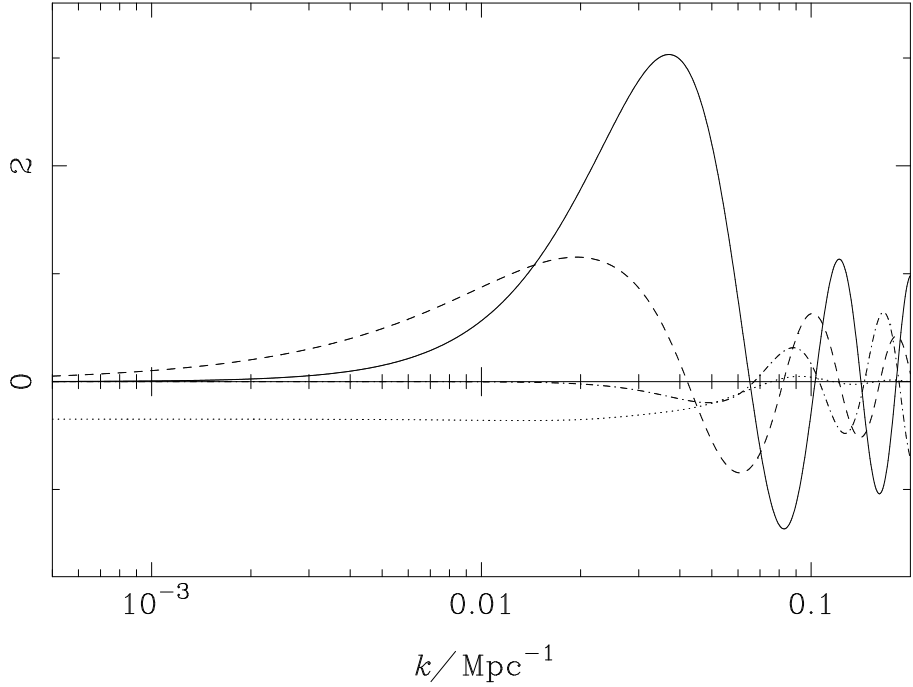}
\end{center}
\caption{{\small Approximate contributions to the CMB temperature anisotropy
from the conditions on the last scattering surface ($z\simeq 1050$)
in a $K=0$ universe with $\Omega_{b}=1$, and
$H_{0}=50\mbox{$\mathrm{kms^{-1}\,Mpc^{-1}}$}$. Only
adiabatic scalar perturbations are considered.
On large scales the usual Sachs-Wolfe term (dotted line)
$- \Phi_{k}/3$ dominates. For adiabatic initial conditions,
the effect of entropy perturbations (dashed-dotted line)
is only important on small scales where the tight-coupling
approximation ceases to hold and adiabaticity is broken. The
coherent sum of the entropy perturbations and the term
$2{\mathcal{H}}_{k}^{-2} \Phi_{k}/9$ (solid line),
as well as the ``dipole'' term
$2{\mathcal{H}}_{k}^{-1}(\Phi_{k} + H^{-1}\dot{\Phi}_{k})/3$
(dashed line) determine the structure of the Doppler peaks.}}
\label{fig_split}
\end{figure}

In Fig.~\ref{fig_split} we plot the individual contributions to the
CMB anisotropy in the baryon dominated limit from
equation~(\ref{eq_deltatbar}).
On large scales (small $k$), the dominant contribution is from the usual
Sachs-Wolfe term $-\Phi_{k}/3$. The effect of entropy perturbations is
negligible on large scales since the baryon-photon fluid is still
tightly-coupled at last scattering on these scales, and our initial conditions
are adiabatic. However, on small scales the entropy perturbations add
coherently with the term $2{\mathcal{H}}_{k}^{-2}\Phi_{k}/9$ to reduce
the ``monopole''
contribution to the anisotropy. These terms along with the ``dipole'' term
determine the structure of the Doppler peaks.

We introduced ${\mathcal{X}}_{k}^{(b)}$ into Eq.~(\ref{eq_monapprox}) to show
explicitly the contribution from entropy perturbations at last scattering,
and to facilitate comparison with other results in the literature.
However, this decomposition into an entropy perturbation term and terms
involving the potential $\Phi_{k}$ is rather unnatural, and
can be replaced by an expression which allows a
more physical interpretation of the temperature anisotropy on
intermediate and small scales.
We use Eq.~(\ref{eq_econshar}) to eliminate ${\mathcal{X}}_{k}^{(b)}$ and
choose $u^{a}$ to coincide with the baryon velocity ($v_{a}^{(b)}=0$;
if the universe is not baryon dominated at recombination, then the
energy-frame should be employed instead of the baryon-frame):
\begin{equation}
{\textstyle \frac{1}{4}} {\mathcal{X}}_{k}^{(\gamma)} +
{\textstyle \frac{S}{k}} (\dot{\sigma}_{k}-w_{k})
\rightarrow {\textstyle \frac{1}{4}} \bar{{\mathcal{X}}}_{k}^{(\gamma)} -
{\textstyle \frac{1}{3}}
\Phi_{k} - {\textstyle \frac{2}{\kappa\rho}}
{\textstyle \frac{K}{S^{2}}}
\Phi_{k} +{\textstyle \frac{2H}{\kappa\rho}} 
\dot{\Phi}_{k},
\end{equation}
where $\bar{{\mathcal{X}}}_{k}^{(\gamma)}$
is the harmonic expansion coefficient of the spatial gradient of the
photon energy density in the energy-frame. This result actually holds
in the energy-frame of any model that is matter dominated at last scattering,
and so includes standard CDM and variants.
We see that in a $K=0$ universe, which is matter dominated at recombination,
the ``monopole'' contribution to
the anisotropy consists of the Sachs-Wolfe term, $-\Phi_{k}Q^{(k)} /3$, which
is dominant on large scales, a term
$\bar{{\mathcal{X}}}_{k}^{(\gamma)}Q^{(k)}/4$ arising from inhomogeneity of the
radiation temperature in the energy-frame, which is important on
intermediate and small scales, and the small term
$2 H^{-1} \dot{\Phi}_{k} Q^{(k)} /3$ arising from non-stationarity of the
potential at last scattering. Since the last scattering surface is
well approximated by a surface of constant radiation temperature, the
spatial gradient of the radiation energy density, in the energy frame,
across the last scattering surface describes a distortion of the last
scattering surface relative to the surfaces of simultaneity of the energy-frame
(which are defined by the average motion of all matter in the universe).
The distortion of the last scattering surface causes photons to incur
extra redshift, due to the expansion of the universe, during propagation from
last scattering to the point of observation.

Finally, if  we consider initially adiabatic perturbations in a
$K=0$ baryon dominated universe on large enough scales that entropy
perturbations may be neglected, we may replace
$\bar{{\mathcal{X}}}_{k}^{(\gamma)}/4$ by the spatial gradient of the
baryon energy
density in the baryon-frame, $\bar{{\mathcal{X}}}_{k}^{(b)}/3$.
Replacing $\Phi_{k}$ with $3{\mathcal{H}}_{k}^{2}
\bar{{\mathcal{X}}}_{k}^{(b)}/2$
(from the constraint~(\ref{eq_econshar}) with $K=0$) and neglecting terms
involving $\dot{\Phi}_{k}$, we obtain the covariant equivalent of Panek's
result~\cite{panek86}:
\begin{equation}
\left(\delta_{T}(e)\right)_{R} = \sum_{k} \left({\textstyle \frac{1}{3}}
\bar{{\mathcal{X}}}_{k}^{(b)} \left[1 - {\textstyle \frac{3}{2}}
{\mathcal{H}}_{k}^{2} \right] Q^{(k)}
\right)_{A} + \sum_{k} \left(\bar{{\mathcal{X}}}_{k}^{(b)}{\mathcal{H}}_{k}
e^{a} Q^{(k)}_{a} \right)_{A}.
\end{equation}

\section{Comparison with the Sachs-Wolfe Result}
\label{sec_sw}

It is instructive to compare the preceding analysis with the
original calculation of Sachs and
Wolfe~\cite{sachs67}. Assuming that the
radiation is nearly isotropic in the frame of the baryons at last
scattering, the temperature of the radiation at $R$ (in the baryon-frame)
is given in terms of the
radiation temperature at the point $A$ in the last scattering surface as
\begin{equation}
T_{R} = \frac{T_{A}}{1+z},
\label{bar_eq16}
\end{equation}
where the redshift $z$ along the null geodesic connecting $A$ and $R$
is given by~\cite{sachs67}
\begin{equation}
1+ z = {\frac{S_{R}}{S_{A}}} \left[1 - \frac{1}{2}
\int_{\eta_{A}}^{\eta_{R}}
\left( \partial_{\eta} h_{ij} e^{i}e^{j} + 2 \partial_{\eta} h_{i0}
e^{i} \right) \, d\eta \right],
\label{bar_eq17}
\end{equation}
where $S_{R}\equiv S(\eta_{R})$ is the background scale factor at $R$ and
similarly for $S_{A}$, $h_{\mu \nu}$ is the metric perturbation in the comoving
gauge (with $h_{00}=0$), $\eta$ is conformal time, and $e^{i}$ $(i=1,2,3)$ is
the direction of photon propagation in the background model. Differencing
the temperature between two directions on the sky we find
\begin{equation}
\left(\frac{\Delta T}{T}\right)_{R} = \left(\frac{\Delta T}{T}\right)_{E}
+\left(\frac{\Delta S}{S}\right)_{E} + \frac{1}{2}
\Delta \left[  \int_{\eta_{E}}^{\eta_{R}}
\left( \partial_{\eta} h_{ij} e^{i}e^{j} + 2 \partial_{\eta} h_{i0}
e^{i} \right) \, d\eta \right],
\label{bar_eq18}
\end{equation}
where $(\Delta M)_{E} \equiv (M)_{A} - (M)_{B}$. The $(\Delta S)_{E}$
term on the right-hand side of Eq.~(\ref{bar_eq18}) is gauge-dependent
(first-order gauge transformations of the form
$\eta \mapsto \eta + f(x^{i})/S(\eta)$,
which preserve the gauge-conditions implicit in Eq.~(\ref{bar_eq17}), move the
field $S$ around in the real universe), while the term $(\Delta T)_{E}$
is gauge-invariant. It follows that the sum of the first two terms on the
right-hand side of Eq.~(\ref{bar_eq18}) is also gauge-dependent.
(A compensating gauge-dependence in the integral term preserves the necessary
gauge-invariance of $(\Delta T)_{R}$.) This sum may be expressed in terms of
the gauge-dependent photon density perturbation $\delta^{(\gamma)}$, defined by
\begin{equation}
\frac{1}{4} \delta^{(\gamma)} \equiv \frac{ T - T^{(0)}}{T^{(0)}},
\label{bar_eq19}
\end{equation}
where $T^{(0)} \equiv T^{(0)}(\eta)$ is the background radiation temperature.
Using the constancy of $T^{(0)} S$, we find
\begin{equation}
\left(\frac{\Delta T}{T}\right)_{E} + \left(\frac{\Delta S}{S}\right)_{E} =
\frac{1}{4} \Delta(\delta^{(\gamma)})_{E}.
\label{bar_eq20}
\end{equation}
The significance of this result is that it is a first-order quantity.
To evaluate $\Delta (\delta^{(\gamma)})_{E}$ to first-order, we need only
locate the last scattering surface to zero-order, since locating the
last scattering surface correctly amounts to a first-order displacement
in a first-order quantity. The implicit choice made by many authors is
to evaluate Eq.~(\ref{bar_eq20}) on the background last scattering surface
(over which $\eta$ is constant). With this choice $\Delta S$ and
$\Delta T^{(0)}$ are both equal to zero, and trivially we obtain the result
$\Delta(\delta^{(\gamma)})_{E}$ evaluated on the background last scattering
surface. Note that Sachs and Wolfe~\cite{sachs67} explicitly made this
choice, although they also made the (gauge-dependent) assumption that the
radiation temperature was constant on the background last scattering surface.
Although it is certainly necessary to locate last scattering
correctly to calculate $\Delta (T)_{E}$ and $\Delta (S)_{E}$ separately,
this is not true for the difference $\Delta (TS)_{E}$, which is all
that is required for CMB calculations.
Physically, this effect arises from the compensation effect: the extra
redshift along a given direction due to the difference between the
scale factor on the true last scattering surface and at the point of
intersection of the geodesic with the background (zero-order)
last scattering surface cancels out the difference in temperature between
the same points on the real and true last scattering surface.
Note that such problems do not arise in the kinetic theory calculation of
Section~\ref{sec-kin}, since the optical depth (which determines the
position of the last scattering surface) multiplies only first-order variables,
and so is needed only to zero-order.

The term $\Delta(\delta^{(\gamma)})_{E}/4$ is often omitted from Sachs-Wolfe
type analyses (for example,~\cite{padmanab-struct}), effectively being
absorbed into a (direction-dependent) ``expected'' temperature,
$\bar{T} \equiv T_{A} S_{A}/S_{R}$ (see~\cite{stoeger95a,panek86}
for alternative choices).
Along with the observed temperature $T_{R}$, this defines a first-order
temperature variation $\delta T/T$ by
\begin{equation}
\frac{\delta T}{T} \equiv \frac{T_{R} - \bar{T}}{\bar{T}}.
\label{bar_eq21}
\end{equation}
This should not be confused with the gauge-invariant temperature fluctuation
from the mean $\delta_{T}(e)$, used in the rest of this paper.
The quantity $\delta T/T$ is gauge-dependent through the gauge-arbitrariness
in the scale factor $S$. Differencing between two directions on the sky, one
finds
\begin{equation}
\Delta \left(\delta T/T\right)_{R} = 
\Delta \left[  \int_{\eta_{E}}^{\eta_{R}}
\left( \partial_{\eta} h_{ij} e^{i}e^{j} + 2 \partial_{\eta} h_{i0}
e^{i} \right) \, d\eta \right],
\label{bar_eq22}
\end{equation}
Of course, this result is entirely equivalent to Eq.~(\ref{bar_eq18}), one
needs only the result
\begin{equation}
\Delta\left(\frac{\delta T}{T}\right)_{E} = \left(
\frac{\Delta T}{T}\right)_{E} - \frac{\Delta (TS)_{E}}{(TS)_{E}},
\label{bar_eq23}
\end{equation}
which follows from the definition of $\bar{T}$. This appears to be the step
where confusion sometimes arises in Sachs-Wolfe type analyses, with authors
comparing the gauge-dependent $\delta T/T$ with the results of observation,
instead of the physically relevant $(\Delta T/T)_{R}$ (or $\delta_{T}(e)$
which is easily derived from the former). As noted in~\cite{stoeger95a},
it is not enough just to difference $\delta T/T$; in general, the result will
not be gauge-invariant and will not include the contribution from
$\delta^{(\gamma)}$ at last scattering.

Finally, we demonstrate how the gauge-invariant contribution
$\bar{{\mathcal{X}}}_{k}^{(\gamma)} Q^{(k)}/4$ to the anisotropy arises in the
Sachs-Wolfe calculation. If we
approximate the (vorticity-free) baryon motion as being geodesic through
recombination to the present, we can choose a comoving gauge with
$h_{i0}=0$ (the synchronous gauge, in which the surfaces of constant
$\eta$ are surfaces of simultaneity for the baryons).
The result given in Eq.~(\ref{bar_eq18}) (with $h_{i0}=0$) still
holds in this gauge, and the conformal time $\eta$ satisfies
$\nabla_{a} \eta = u_{a}^{(b)}/S(\eta)$, where $u_{a}^{(b)}$ is the baryon
velocity. It follows that the background scale factor satisfies
${}^{(3)}\nabla_{a} S = 0$ and $\dot{S} = \partial_{\eta} S / S = \theta S /3+
{\mathcal{O}}(1)$,
where $\theta = \nabla^{a} u_{a}^{(b)}$ is the covariant expansion of the
baryon-frame. With this gauge-condition,
we have removed the gauge-freedom to
perform the transformation $\eta \mapsto \eta + f(x^{i})/S$ (the
gauge-condition ${}^{(3)}\nabla_{a} \eta = 0$ forces $f(x^{i})$ to be a
constant),
with the result that $\Delta S/S$ is gauge-invariant.
Under these conditions, we may express the first two
terms on the right-hand side of Eq.~(\ref{bar_eq18}) in
terms of covariant, gauge-invariant variables as follows:
\begin{align}
\left(\frac{\Delta T}{T}\right)_{E} +
\left(\frac{\Delta S}{S}\right)_{E} & = 
\frac{1}{4} \int_{B}^{A} \left(\frac{\nabla_{a}\rho^{(\gamma)}}
{\rho^{(\gamma)}}
+ 4 \frac{\nabla_{a}S}{S} \right) dx^{a} \nonumber \\
& =  \frac{1}{4} \int_{B}^{A} \left(\frac{\nabla_{a}
\rho^{(\gamma)}}{\rho^{(\gamma)}}
+ \frac{4}{3} \theta u_{a}^{(b)} \right) dx^{a} \nonumber \\
& = \sum_{k} \Delta\left({\textstyle \frac{1}{4}}
\bar{{\mathcal{X}}}_{k}^{(\gamma)}Q^{(k)} 
\right)_{E},
\label{bar_eq24}
\end{align}
where $A$ and $B$ are the points of intersection of the two geodesics through
the observation point $R$ with the last scattering surface, $dx^{a}$ lies in
the last scattering surface ($dx^{a}u_{a}^{(b)}={\mathcal{O}}(1)$),
and we have
replaced $3\dot{S}/S$ by the covariantly defined $\theta$ which is correct to
the required order. In this manner, we recover the
$\bar{{\mathcal{X}}}_{k}^{(\gamma)}Q^{(k)} /4$ contribution to the temperature
anisotropy.

\section{Conclusion}

Starting from a covariant and gauge-invariant formulation of the Boltzmann
equation, we have derived a new expression for the CMB temperature anisotropy
under the instantaneous recombination approximation, valid for scalar
perturbations in open, closed and flat universes. Our expression uses only
covariantly-defined variables, and is manifestly gauge-invariant.
The result is more useful in multicomponent models with scalar perturbations
than earlier covariant results~\cite{ellis-er97,LC97-er,dunsby96b}.
In the case of a universe which is baryon-dominated at
recombination, we find a simple expression for the anisotropy which corrects
a similar result by Dunsby~\cite{dunsby96b}. By making use of numerical
solutions to the perturbation equations, we have discussed the conditions on
the last scattering surface and their contributions to the characteristic
features of the CMB power spectrum.
We ended with a discussion of the original Sachs-Wolfe calculation for the
temperature anisotropy. We have discussed why it is not necessary to locate
accurately the last scattering surface in such calculations (because of the
compensation effect), and how the extra term in the Sachs-Wolfe calculation,
reported recently by Ellis and Dunsby~\cite{ellis-er97}, is missed in
many calculations which employ a gauge-dependent ``expected temperature'',
since the angular dependence of this temperature is often overlooked.
For a universe which is matter dominated at recombination, but not
necessarily adiabatic, the extra term is the spatial gradient of the
radiation energy density in the energy-frame,
$(\bar{{\mathcal{X}}}_{k}^{(\gamma)} Q^{(k)} /4)_{A}$. For models with isothermal
surfaces of last scattering, this
inhomogeneity describes a distortion of the last scattering surface
relative to the surfaces of simultaneity of the energy-frame. The extra
redshift incurred by this distortion is a significant component of the
temperature anisotropy on intermediate and small scales.


\end{document}